\documentclass[sigconf,edbt]{acmart-edbt2021}
\usepackage[ruled]{algorithm2e}
\usepackage{amsmath}
\setcopyright{rightsretained}
\newcommand{\ie}{\textit{i}.\textit{e}., }
\newcommand{\eg}{\textit{e}.\textit{g}., }

\newcommand{\oli}[1]{\textcolor{black}{#1}}

\acmDOI{}

\acmISBN{978-3-89318-084-4}

\acmConference[EDBT 2021]{24th International Conference on Extending Database Technology (EDBT)}{March 23-26, 2021}{Nicosia, Cyprus} 
\acmYear{2021}

\begin{document}

\title{Knowledge Graph Management on the Edge}

\author{Weiqin Xu}

\affiliation{%
  \institution{LIGM Univ Paris Est Marne la Vall\'ee, CNRS, France}
  \city{Engie Lab, CRIGEN, Stains, France} 
}
\email{weiqin.xu@u-pem.fr}

\author{Olivier Cur\'e}

\affiliation{%
  \institution{LIGM Univ Paris Est Marne la Vall\'ee, CNRS, France}
}
\email{olivier.cure@u-pem.fr}

\author{Philippe Calvez}

\affiliation{%
  \institution{Engie Lab, CRIGEN}
  \city{Stains} 
  \state{France} 
}
\email{philippe.calvez1@engie.com}

\renewcommand{\shortauthors}{}

\begin{abstract}
Edge computing emerges as an innovative platform for services requiring low latency decision making. Its success partly depends on the existence of efficient data management systems.
We consider that knowledge graph management systems have a key role to play in this context due to their data integration and reasoning features.
In this paper, we present SuccinctEdge, a compact, decompression-free, self-index, in-memory RDF store that can answer SPARQL queries, including those requiring reasoning services associated to some ontology.
We provide details on its design and implementation before demonstrating its efficiency on real-world and synthetic datasets.    
\end{abstract}

\maketitle

\section{Introduction}
Edge computing\cite{edgeComp} corresponds to a processing paradigm that brings storage, management, and processing of huge amounts of data closer to the location where it needs to be performed.
As such, this emerging trend complements a cloud computing approach by supporting the design of highly local context aware and responsive services, hence eliminating round trips to the Cloud, as well as mask cloud computing outages. A key challenge for systems designed for edge computing is an efficient data management in the context of mobile devices and sensors/actuators which generally have stringent requirements on energy consumption as well as memory, CPU usages and network bandwidth.

Our prototype system, SuccinctEdge\footnote{https://github.com/xwq610728213/SuccinctEdge}, has been designed for edge computing from the get go and adopts \oli{the Resource Description Framework (RDF). The adoption of this data model is  motivated by the data integration and reasoning facilities it provides. Considering the former, the Linked Data principles\footnote{https://www.w3.org/wiki/LinkedData} together with the large set of standardized Knowledge Graphs (KGs) via the Linked open Data initiatives\footnote{https://lod-cloud.net/} ease the design of Internet of Things (IoT) applications. For instance, ontologies such as  the Sensor, Observation, Sample, Actuator (SOSA \footnote{http://www.w3.org/TR/ns/sosa}, Quantities, Units, Dimensions, and Types (QUDT) \footnote{http://qudt.org/schema/qudt} or Smart Applicances Reference (SAREF)\footnote{https://ontology.tno.nl/saref.ttl} are considerably simplify the task of describing, manipulating and connecting sensors and actuators. These ontologies are also quite useful for the smart management of the measures in a precise context where it may be necessary to infer implicit consequences from explicitly represented knowledge.}

SuccintEdge favors a compressed, single index storage approach to a solution based on multiple indexes that could potentially improve query execution but at the cost of a higher memory footprint. The applications we are targeting with SuccinctEdge are the processing of a flow of RDF graphs (sent from sensors or actuators) which are sharing a common topology. These are graphs being continuously queried by a set of SPARQL queries. A typical use case is that these queries are searching for anomalies that are occurring over a network of sensors (see Section {sec:example} for a motivating example). As a result, these queries are executed once per graph instance.

Our system makes an intensive use of succinct data structures (SDS)\cite{DBLP:conf/cpm/Navarro12}, a family of data structures that adopts a compression rate close to theoretical optimum, but simultaneously allows efficient decompression-free query operations on the compressed data. Together with our single index approach, SDS guarantees a low memory footprint that fits with an in-memory storage approach. The decompression-free aspects also tends to reduce the number of CPU cycles on standard queries and inferences.

SuccinctEdge's reasoning services are based on the LiteMat encoding solution\cite{DBLP:conf/esws/CureXNC19}. This approach prevents inference materialization and reduces the cost of the SPARQL query rewriting task, the two most frequent reasoning solutions in RDF stores. As a result of encoding most triple entries with integer values, this approach, like most RDF Stores, compresses the size of RDF stores and thus limits the memory footprint of a given graph.

SuccinctEdge is addressing the compact storage and efficient querying of RDF data via SPARQL queries in the presence of RDFS reasoning in an edge computing environment. The main contributions of this paper are to (i) present a self-index, compact, in-memory storage layout based on the bitmap and wavelet tree SDSs, (ii) propose a decompression-free (\ie the SDS compressed graph does not need any decompression step to enable query processing), efficient query processing and optimization of SPARQL basic graph patterns which are transformed into access, rank and select SDS operations, (iii) support reasoning during query processing using a smart encoding approach and (iv) propose a simple and automatic approach to express complex queries requiring inferences by preventing 
end-users to learn the details of used ontologies and ontology annotations used at each sensor. 

We demonstrate the efficiency of our implementation on an evaluation conducted on real-world and synthetic datasets.
This paper is organized as follows. In Section \ref{sec:example}, we motivate our approach with a real-world example in an industrial setting. In Section \ref{sec:background}, we provide some background knowledge. Section \ref{sec:archi} presents the overall architecture of SuccinctEdge. The query optimizer and processor is presented in Section \ref{sec:query}. Section \ref{sec:related} relates our research to existing systems and Section \ref{sec:eval} provides a detailed experimentation conducted on real-world and synthetic data. We conclude the paper and present directions for future work in Section \ref{sec:conclusion}.

\section{Motivating example}
\label{sec:example}

In this paper, we consider an upcoming deployment of SuccinctEdge at some of ENGIE's buildings where an IoT network is deployed. 
ENGIE is a multinational company operating in fields such as energy transition, generation and distribution. 

Our running example focuses on data harvested from a building management system with a first focus on potable water distribution. Intuitively, a flow of measures are obtained from a network of sensors. A thorough analysis permits to detect anomalies such as leaks or other abnormal situations from, for instance, pressure and flow measurements. The measures are usually represented as text files (\eg CSV) but, thanks to some mapping assertions and dedicated digital services deployed through APIs, are transformed into a form of RDF graph (to be detailed later in this paper) and annotated with concepts of a domain ontology. 

\begin{figure*}[ht]
\centering
\includegraphics[scale=0.4]{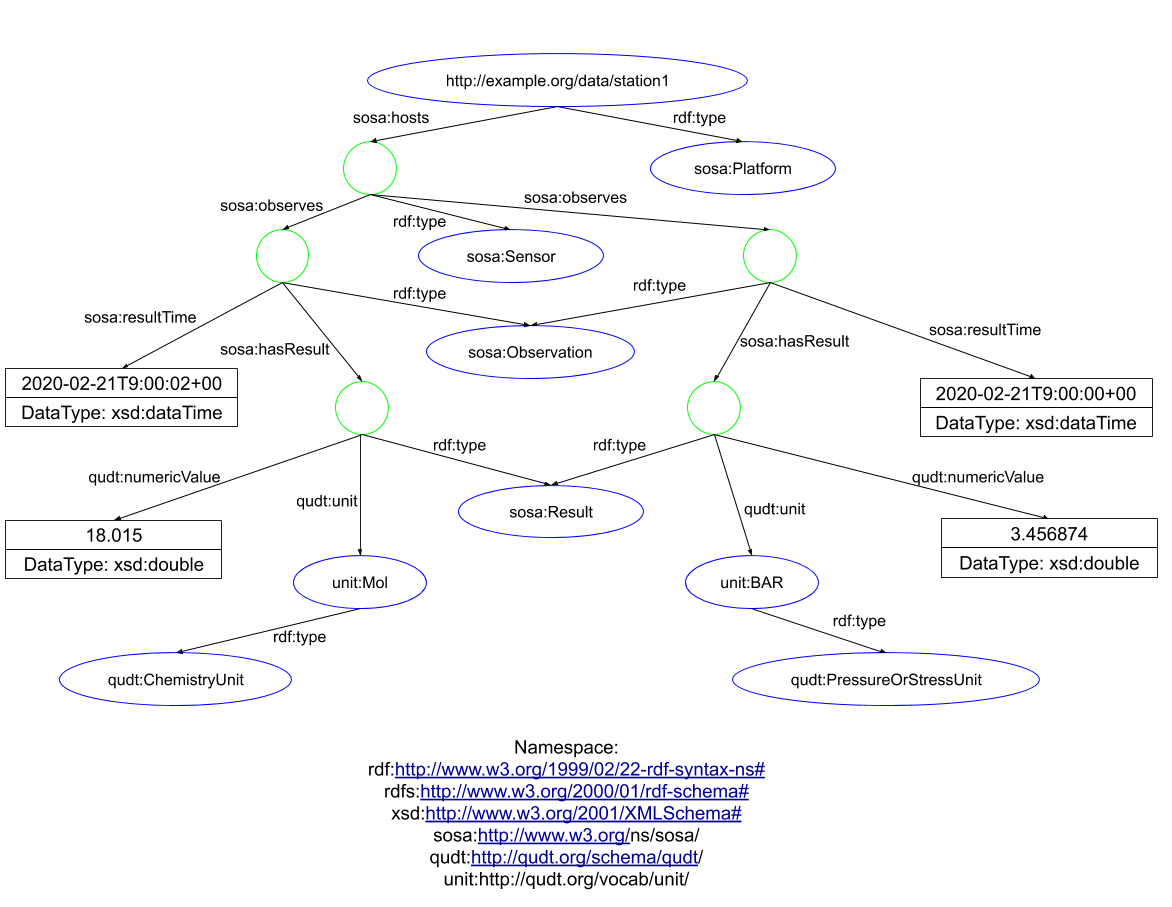}
\caption{Graph extract of our use-case (green nodes are blank nodes)}
\label{fig:graphExtract}
\end{figure*}

Figure \ref{fig:graphExtract} proposes an extract of such a graph which concerns pressure and chemistry measures related to the water distribution management. Given such graph instances, our SuccinctEdge system executes queries that can detect some patterns such as anomalies linked to the water management system, \eg incorrect chemistry properties, network leak, etc. In a non edge computing context, each measure would transit on a computing network to a more powerful machine that could process the anomaly detection. Such an approach as several drawbacks: (i) it makes an intensive use of the computing and communication network which can rapidly be overloaded, \eg devices on the edge of the network generally have low bandwidth,
(ii) the high-end computing machine also risks to be overcharged and stressed from the amount of data received (potentially from hundreds to thousands of sensors) and 
(iii) sending these data packets over the network is not cost-free for these sensors, \eg in terms of energy consumption.

In a context where anomalies are the exception, it makes sense to detect anomalies as close as possible to the sensors since it would require to (i) send fewer data over the computing network as that would occur only in anomaly cases, (ii) reduce decision latency and (iii) keep the high-end computing machine unstressed.

In the context of our experimentation at ENGIE, we are designing a query-based anomaly detection approach that does not require from the end-users a high level of expertise on the underlying domain ontology and its reasoning services. Hence these users only express queries in relatively high concept terms and do not have to worry about the inferences which are handled automatically by the system. 
Expressing a query with abstract concepts, \ie high in the concept hierarchy, permits to write a single query that can tackle sensors performing similar measures but annotated with different concepts and possibly with different measure units. This is an important requisite for our use case where different sensor brands and types can coexist in a given network. The simplicity of this approach was highly expected from ENGIE for productivity reasons. In fact, it enables its sensor personnel to concentrate on their tasks and not on adapting a given query to the potentially large number of sensors in an industrial setting.
For instance, in the following real-world example, 2 sensor platforms are measuring similar values, \eg pressure and chemistry-related, but each sensor annotates them with different concepts. Considering Station1 the pressure and chemistry are respectively annotated with $qudt:PressureOrStressUnit$ and $qudt:Chemistry$, while for Station2, it is resp. $qudt:Pressure$ and $qudt:AmountOfSubstanceUnit$. Moreover, the pressure value in Station1 is expressed in Bar while it is measured in hecto Pascal in Station2.

Since, the QUDT ontology\footnote{https://qudt.org/} states that:

$qudt:AmountOfSubstanceUnit \sqsubseteq qudt:Chemistry \sqsubseteq qudt:ScienceUnit$ and $qudt:PressureOrStressUnit \sqsubseteq qudt:PressureUnit \sqsubseteq qudt:MechanicsUnit$, a single SPARQL query can be written to address the specificities of each sensor at these 2 stations. The following query detects anomalies related to an incorrect pressure value (either expressed in Bar or HectoPascal) for sensors of stations 1 and 2:

\begin{small}
\begin{verbatim}
SELECT ?x ?s ?ts ?v1 WHERE {
?x a sosa:Platform; sosa:hosts ?s. 
?s sosa:observes ?o; a sosa:Sensor. 
?o sosa:hasResult ?y; a sosa:Observation; 
sosa:resultTime ?ts. ?y a sosa:Result; 
qudt:numericValue ?v1; qudt:unit ?u1. 
?u1 a qudt:PressureUnit. FILTER (?newV<3.00 || ?newV>4.50)
BIND(if(regex(str(?u1),"http://qudt.org/vocab/unit/BAR"),?v1,
if(regex(str(?u1),"http://qudt.org/vocab/unit/HectoPA")
,?v1/1000,0)) as ?newV) }
\end{verbatim}
\end{small}

\section{Background Knowledge}
\label{sec:background}
\subsection{Semantic Web standards}
RDF is the W3C recommendation schema-free data model that supports the description of data on the Web. It takes the form of a graph consisting  of a set of triples. Each triple is  composed of a subject, a predicate and an object. Properties can be qualified as object or datatype. They both related a URI (or blank node) to respectively a URI or a literal.
SPARQL, another W3C recommendation, enables to express queries over RDF data. The syntax is inspired by SQL's SELECT-FROM-WHERE but it uses an approach based on matching a basic graph pattern (BGP), \ie a set of triple patterns (TP), on an RDF graph to retrieve query answer sets.
Finally, RDF Schema (RDFS) and Web Ontology Languages (OWL) enable the description of vocabulary semantics used in RDF datasets. They support inference services based on their respective expressiveness.

\subsection{LiteMat}
\label{s:litemat}
LiteMat is a semantic-aware encoding scheme that compresses RDF data sets and supports reasoning services associated to the RDFS ontology language. In this work, we are focusing on the $\rho$df\cite{Munoz:2009:SEM:1640541.1640852} subset of RDFS, \ie inferences associated to the \texttt{rdfs:domain}, \texttt{rdfs:range}, \texttt{rdfs:subClassOf} and \texttt{rdfs:subPropertyOf} properties. To address inferences drawn from these last two RDFS predicates, we attribute numerical identifiers to ontology terms, \ie concepts and predicates, that are supporting the semantics. 
This is performed by prefixing the encoding of a term with the encoding of its direct parent. 
This approach only works if an encoding is computed using a  binary representation and all binary encoding entries are all of the same length.
The encoding is performed using a top-down approach, \eg starting from the most specific concept of the hierarchy (typically \texttt{owl:Thing}, \texttt{owl:topObjectProperty} and \texttt{owl:topDataProperty} for respectively the concept, object property and datatype property hierarchies), until all leaves are processed. Then a normalization is performed to guarantee that all encoding entries have the same length, \ie by setting right-most bits to 0.

\begin{figure}
\centering
\includegraphics[scale=0.27]{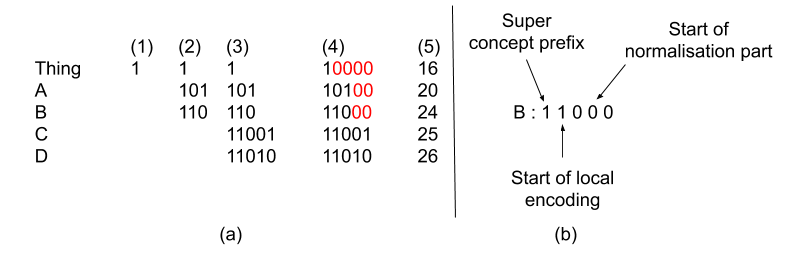}
\caption{LiteMat encoding example}
\label{fig:encoding}
\end{figure}

We now provide an example on a concept hierarchy (a similar approach is used for property hierarchies). In Figure \ref{fig:encoding}, we consider a small ontology extract containing the following axioms: $A \sqsubseteq Thing$, $B \sqsubseteq Thing$, $C \sqsubseteq B$ and $D \sqsubseteq B$. Figure \ref{fig:encoding}(a) highlight the top-down encoding approach with (1) setting the local identifier of $Thing$, (2) its direct sub-concepts ($A$ and $B$) and $B$'s sub-concepts in (3). Then, in (4) the normalization step is performed, \ie added right-most bits are written in red. Column (5) provides the integer value attributed to each concept. 

The mapping between URIs and their identifiers are stored in dictionaries, two for the concepts and two for the properties to support a bidirectional retrieval, \ie from a URI to its identifier and from an identifier to its URI.
Moreover, in the former dictionaries, additional identifier metadata are stored.
For instance, the local length (binary length before the normalization phase) of each dictionary entry is stored along the final identifier entry. Figure \ref{fig:encoding}(b) emphasizes the different metadata of the LiteMat encoding for the $B$ concept: super concept identifier part, start of local encoding and start of the normalization part.

The semantic encoding of concepts and predicates supports reasoning services usually required at query processing time. For instance, consider a query asking for the pressure value of sensors of type \texttt{S1}. This would be expressed as the following two triple patterns: \texttt{?x pressureValue ?v. ?x type S1}. In the case sensor concept \texttt{S1} has n sub-concepts, then a naive query reformulation requires to run the union of n+1 queries. With LiteMat's semantic-aware encoding, we are able, using two bit-shift operations and an addition, to compute the identifier interval, \ie [lowerBound, upperBound), of all direct and indirect sub-concepts of \texttt{S1}.  And thus we can compute this query with a simple reformulation: (i) replacing the concept \texttt{S1} with a new variable : \texttt{?x type ?newVar} and (ii) introducing a filter clause constraining values of this variable: \texttt{FILTER (?newVar$>=$lowerBound \&\& ?newVar$<$upperBound)}.

Considering the instance dictionary, each distinct entry is assigned an arbitrary unique integer value.  

\subsection{Succinct Data structures}
\label{s:sds}

SDS represents a family of data structures that stores data in a compact way, but still allows some efficient data access operations without decompression. There are different types of SDS, among which we consider Wavelet Tree (WT) and BitMap (BM).  SuccinctEdge represents an RDF graph into a combination of these two structures to reach a very compact size without loss of query efficiency. 

BM is the most basic SDS we are using in SuccinctEdge. It is a sequence of bits with some extra information to support the efficient execution of SDS operations. BM is the basic building block of WT's nodes (as each node in the tree is a BM), but it also relates different WTs in SuccinctEdge's triple representation (further details in Section \ref{sec:archi}).

\begin{figure}
\centering
\includegraphics[scale=0.3]{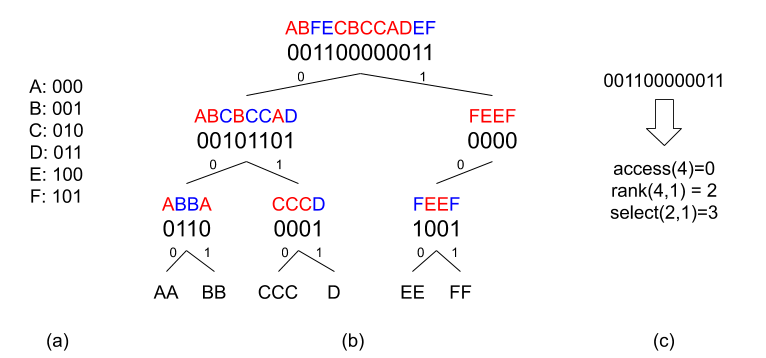}
\caption{Wavelet Tree example with its dictionary}
\label{fig:wt_example}
\end{figure}

WT \cite{DBLP:conf/cpm/Navarro12}, whose name reveals some affinity with the idea of the wavelet packet decomposition in signal processing, refers to a data structure which decomposes a data sequence into a set of nodes of a balanced binary tree. An example of a WT is given in Figure \ref{fig:wt_example}b. Suppose that we have a sequence $ABFECBCCADEF$, where each letter is mapped with an identifier in an incremental order, \eg $A$ is denoted with 0, $B$ is denoted with 1 (see dictionary in \ref{fig:wt_example}a). A tree structure is constructed from this sequence as follows: each level of this tree divides the sequence of previous node into two sub-sequences by the corresponding bit. For example, from root to the first level, $ABFECBCCADEF$ is divided into $ABCBCCAD$ and $FEEF$ by the first bit of each identifier entry. This strategy is applied recursively until each leaf is computed.

SDS support three operations to access data: $Rank$, $Select$ and $Access$. Given a sequence $S$, the operation $S$. $Access(i)$ (also denoted as $S[i]$) refers to the $(i+1)^{th}$ element in $S$. $S.Rank(i,c)$ returns the number of occurrences of $c$ from $S$'s beginning to index $i$. Finally, $S.Select(i,c)$ returns the index of $i^{th}$ occurrence of element $c$ in $S$. These operations can be computed in O(1) for BM and O(log n) for WT where n is the size of the vocabulary. Figure \ref{fig:wt_example}c provides an example over a simple BM. Dedicated algorithms permit to compute these 3 operations over WT. 
In Section \ref{sec:query}, we will present a translation approach from SPARQL BGP to SDS operations.



\section{Architecture overview}
\label{sec:archi}

\begin{figure*}[t]
\centering
\includegraphics[scale=0.15]{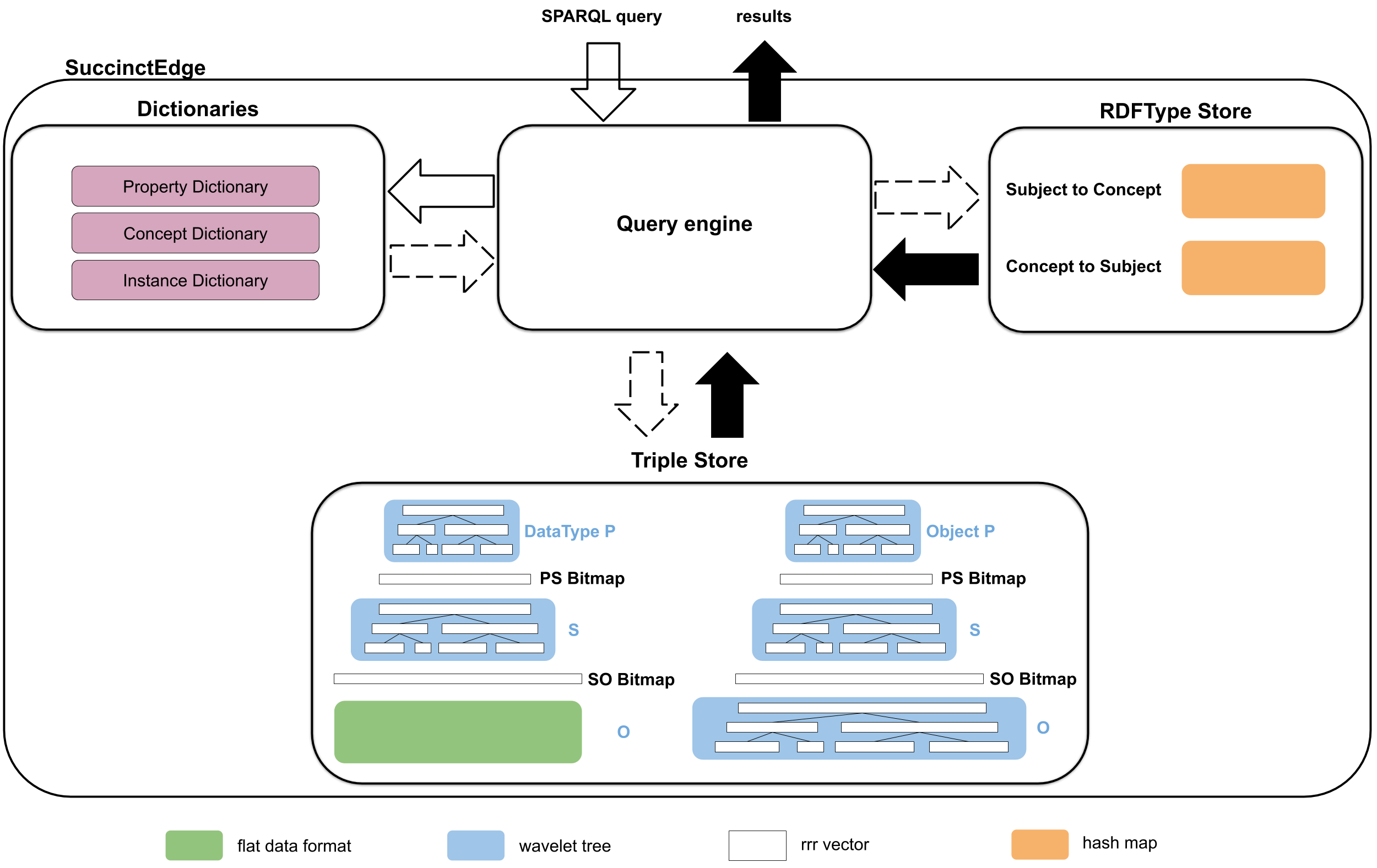}
\caption{Architecture overview of SuccinctEdge}
\label{fig:succinctEdgeArchi}
\end{figure*}

\begin{figure*}
\centering
\includegraphics[scale=0.35]{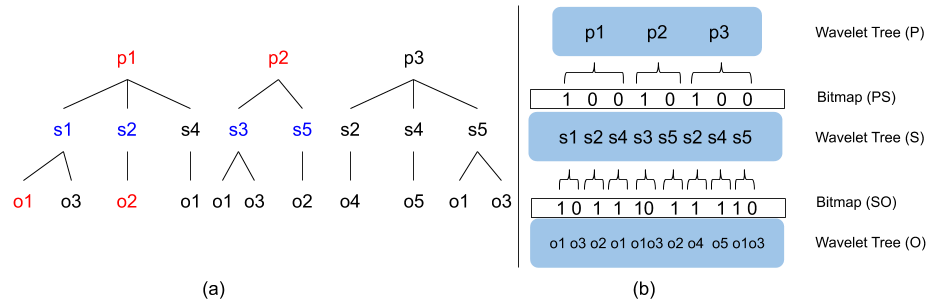}
\caption{RDF graph representation: (a) as a PSO-based forest and (b) in SuccinctEdge as a combination of wavelet trees and bitmaps (only considering object properties) }
\label{fig:triple_store_example}
\end{figure*}

Before providing an overview of the SuccinctEdge RDF store, we describe a standard running setting at an ENGIE building. Typically, the person responsible for the building maintenance supervises a set of IoT devices from a SuccinctEdge server. From this central computer, the administrator is able to register new IoT devices installed in this set of buildings. Each IoT device typically runs a SuccinctEdge instance (client) which can execute many SPARQL queries. The administrator receives alerts from SuccinctEdge instances has abnormal sensor measures are occurring. Hence, each sensor modification (\eg a sensor is replaced due to a failure, a sensor data schema is modified) must go through an administration step which is performed on a central computer. Apart from such maintenance operations, this server also performs the pre-processing task consisting of encoding ontologies using the LiteMat scheme. In this context, and we consider in a large number of industrial settings, the ontologies are stable and thus rarely change. As explained previously, in SuccinctEdge, these ontologies take the form of a set of dictionaries (since their semantics are encoded via the use of LiteMat). These dictionaries are broadcasted to the different SuccinctEdge instances running at the edge. 

An overview of SuccinctEdge's architecture is presented in Figure \ref{fig:succinctEdgeArchi}. Like most RDF stores, all triples are encoded according to some dictionaries. The underlying basic concept of a dictionary is to provide a bijective function mapping long terms (\eg URIs, blank nodes or literals) to short identifiers (\eg integers). More precisely, a dictionary should provide two basics operations : \texttt{string-to-id} and \texttt{id-to-string} (also referred in the literature as \texttt{locate} and \texttt{extract} operations). 
In a typical use of SuccinctEdge, the query engine will call the \texttt{locate} operation to rewrite the query into a list to match the data encoding, while the \texttt{extract} operation will be called to translate the result into the original format. In our case, we are using LiteMat (as presented in Section \ref{s:litemat}) to generate the concept, property and individual dictionaries. 

The Triple store component adopts a single index based on the predicate, subject, object (PSO) triple permutation. That is, the triples of the graph are sorted in ascending order over the P, S and O values of our dictionaries. The PSO order is motivated by the fact that the basic graph pattern of queries submitted to SuccinctEdge have predicates filled in with URIs (as opposed to variables). This corresponds to typical IoT use cases where queries are retrieving information from measures rather than serving to discover patterns in the graphs. In fact, there is no need for discovery since the graph patterns are well known in advance and are very rarely modified (\ie mostly due to sensor failure in industrial use-cases).  

The Triple store component also highlights that we make a distinction between object (expect \texttt{rdf:type}) and datatype properties. In the former, objects are individuals and thus encoded with the respective instance dictionary while in the latter, objects are literals and stored using a flat data structure to store literals. This last data structure is motivated by the fact that it is not reasonable to create an entry in the instance dictionary for each new literal value. Intuitively, a sensor generally sends numerical values corresponding to physical measurement at a given time. Depending on the precision of these measures, the amount of different values to store in the instance dictionary is potentially infinite. So, we prefer to store the values as they have been sent by sensors, possibly with some redundancy, in order to prevent a complex and costly individual dictionary management.

 In terms of data structures, WTs are used for the property and subject layers as well as the object layer for object properties. In order to relate a WT of one layer to another, we are using a BM. Figure \ref{fig:triple_store_example}(b) represents the triple set of Figure \ref{fig:triple_store_example}(a) where a WT corresponds to balanced tree of BMs. Intuitively the PS (respectively SO) bitmap permits to link a given P (resp. S) to several S (resp. O) values. In Figure \ref{fig:triple_store_example}(b), p1 is connected to s1, s2 and s4 because the PS bitmap starts with a 100 sequence: '1' states that the sequence of p1 starts with a given subject (s1) and the '00' states that 2 other subjects are linked to p1. Moreover, the 4th bit in the PS BM (\ie set to '1') starts the sequence of the second property entry in the P WT (\ie p2).

Finally, triples containing a \texttt{rdf:type} property are stored in the RDFType store layout. These triples generally represent an important proportion of the triple set in real-world RDF datasets. We simply store them in a red-black tree in order to maintain the search complexity to O(log(n)) while being fast when we insert \texttt{rdf:type} triples during database construction.

\section{Query processing and optimization}
\label{sec:query}
\oli{In this section, we present the query optimization and processing solutions developed for SuccinctEdge.
Their main goals are respectively to define an efficient TP join ordering, by combining heuristic and cost-based approaches, and to generate a physical plan composed of SDS operations (\ie access, rank and select).}   

\begin{figure*}[h]
\centering
\includegraphics[scale=0.4]{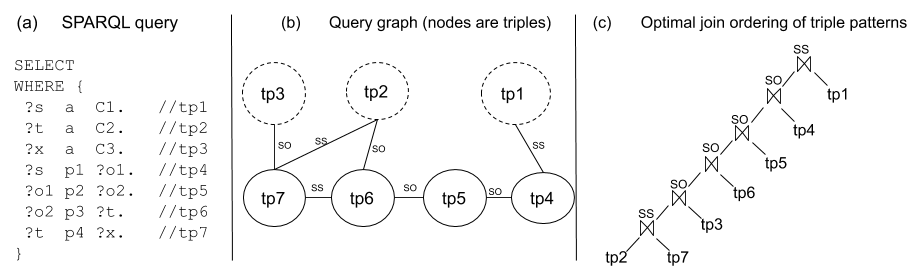}
\caption{Query, query graph and join ordering) }
\label{fig:queryGraph}
\end{figure*}

\subsection{Query Optimization}
\oli{The design of our query optimizer considers the limitations of the devices on  which SuccinctEdge is running on, \ie limited memory space and computing power. Due to these constraints, our system only generates left-deep join trees since they are generally considered to reduce the amount of memory used by the search process.}

\oli{As stated in \cite{DBLP:conf/sigmod/NeumannW09}, join ordering is the most crucial issue in SPARQL query optimization. This is mainly due to the potentially high number of triple patterns and thus of join operations that one can find in BGPs. For instance, in our IoT building management experimentation, we have frequently encountered queries in the range of 10 joins.}

\oli{In order to optimize a given SPARQL query, our query engine constructs a query graph where each TP of the SPARQL query corresponds to a node of the query graph. 
Each query graph node is also annotated to state whether its property is rdf:type or not.  
The nodes in this graph are connected if they share a common variable, hence forming a join. Moreover, the edges of this query graph are labelled with a join type, either SO or SS for respectively subject-object and subject-subject joins.}

\begin{example}
\oli{Figure \ref{fig:queryGraph}(b) displays the query graph associated with the SPARQL query presented in Figure \ref{fig:queryGraph}(a). This query contains 7 TPs, denoted tp1 .. tp7. The dotted nodes in the query graph correspond to rdf:type TPs.}
\end{example}

\oli{Given a query graph, our optimizer uses Algorithm \ref{algo_orderTP} to produce a join order. Intuitively, starting from a given TP, it invokes an overloaded $getMostSelective$ method to search for the next TP to join with. This method uses a set of static rules together with some data statistics. In terms of the former, we have been influenced by Heuristic 1 of \cite{DBLP:conf/edbt/TsialiamanisSFCB12} which defines an execution order for the 8 possible TP combinations. In the context of SuccinctEdge, we do not need to consider all combinations since, as stated previously, triple patterns with either zero or three variables, \ie $(s,p,o)$ and ($?s,?p,?o)$, are highly unlikely to occur in a real-world IoT SPARQL query.}
\oli{Intuitively, this heuristic states that TPs with the fewest variables should be executed first. Our adaptation re-orders the original proposition by taking into account the fact that our access paths are limited to PSO for non \texttt{rdf:type} properties and to SO/OS paths for \texttt{rdf:type} triples}. 
\oli{As presented in Section \ref{sec:archi}, the latter access path (SO/OS on \texttt{rdf:type}) is more efficient than the one based on the SDS structures. We can thus define the following TP order: }

\oli{(s,rdf:type,?o)>(?s,rdf:type,o)>(s,p,?o)>(?s,p,o)>(?s,p,?o), where p denotes any property different from rdf:type and the relation tp1 > tp2 states that tp1 should be executed before tp2. The (s,p,?o)>(?s,p,o) order is due to the navigation mode in our multi-layer SDS triple representation which is PSO based, \ie it is more efficient to retrieve objects given a subject/property pair than to compute subjects given an object/property pair. Finally, we do not consider that the (?s rdf:type ?o) TP is relevant in a practical IoT context.}

\oli{This first heuristic is generally not sufficient to decide which TP to execute among a set of other TPs. Hence, we are considering a second heuristic that takes into consideration the linearity required by a left-deep join tree and examines the types of join possible between several TPs. Due to the PSO self-index SDS structure used for non-rdf:type triples, SS joins are preferred over SO joins, ie $S \bowtie S > S \bowtie O$. Other forms of joins, \ie SP, OP, PP have a lower priority since they are rarely encountered in the setting where SuccinctEdge is relevant.}

\oli{In order to minimize intermediate results, the optimizer also relies on a set of statistics computed at dictionary creation-time. Intuitively, each dictionary persists the number of occurrences of each of its entries, \ie concept, property and non-literal individuals. Our statistic approach considers the hierarchy position of a given concept or property when computing the total number of triples its involved in. For instance, with the following concept hierarchy $C_2 \sqsubseteq C_1 \sqsubseteq C_0$ and $C_3 \sqsubseteq C_0$, the total number of triples involving of $C_0$ will be the sum of triples involving $C_i$ with i $\in (0,1,2,3)$. A similar process is applied to get the correct statistics for properties involved in a property hierarchy. Finally, some statistics are also computed at run-time, \eg the BM and WT data structures facilitate the computation of certain statistics. For instance, Algorithm \ref{algo_cal_p_number} computes the number of triples containing a certain property.}

\oli{Algorithm \ref{algo_orderTP} first starts with the identification of  the most selective rdf:type TP with an SS join. In the case it does not find an rdf:type TP or finds only rdf:type TP connected with SO joins, it then selects a non-rdf:type TP to start with. In the case several TPs satisfy our constraint, the statistics permit to take a decision. That first TP is appended to our tpOrder sequence. We then loop over the remaining nodes of the query graph until all TPs have been added to the sequence. At each iteration of the loop, the $getMostSelective$ method considers TPs in the tpOrder sequence and searches for the next TP to append to this sequence. This search is again based on our two heuristics and the usage of statistics. }

\begin{example}
\oli{
The left-deep join tree displayed in Figure \ref{fig:queryGraph}(c) has been defined using Algorithm \ref{algo_orderTP} considering that tp2 is more selective than tp1, \ie the number of occurrences of C2 is lower than the one C1. Once tp2 has been selected, the optimizer has the choice to join it with tp6 or tp7. tp7 is chosen since a SS join is preferred to a SO join. At this stage, the number of occurrences of concept C3, \ie tp3, can be lower than the number of already computed  binding for ?x, and thus tp3 is selected. Given that tp2, tp7 and tp3 have already been considered, tp6 is the only alternative that can be considered and similarly for the remaining TPs, \ie tp5, tp4 and tp1.
}
\end{example}

\begin{algorithm}
    \caption{Computation of a TP order}
    \label{algo_orderTP}
    \LinesNumbered
    \KwIn{query graph G}
    \KwOut{ordered sequence of TPs}
    \BlankLine
    $tpOrder\leftarrow \emptyset$\;
    $n\leftarrow getMostSelective(rdf:type)$\;
    $tpOrder \leftarrow tpOrder + n$\;
    \While{not all G nodes are in tpOrder}{
        $n\leftarrow getMostSelective(tpOrder)$\;
        $tpOrder \leftarrow tpOrder + n$\;
    }
    return tpOrder\;
\end{algorithm}

\begin{algorithm}
   \caption{Compute the number of triples corresponding to a certain predicate.}
    \label{algo_cal_p_number}
     \LinesNumbered
     \KwIn{Predicate p}
     \KwOut{Number n}
     \BlankLine
     $id_p\leftarrow FindIdFromDictionary(p)$\;
     $index_p\leftarrow wt_p.select(1,id_p)$\;
     $index_{sBegin}\leftarrow
     bitmap_{ps}.select(index_p + 1, 1)$\;
     $index_{sEnd}\leftarrow
     bitmap_{ps}.select(index_p + 2, 1)$\;
     $index_{oBegin}\leftarrow
     bitmap_{so}.select(index_{sBegin} + 1, 1)$\;
     $index_{oEnd}\leftarrow
     bitmap_{so}.select(index_{sEnd} + 2, 1)$\;
     $n\leftarrow index_{oEnd}-index_{oBegin}$\;
     return n\;
\end{algorithm}

\subsection{Query processing}
\oli{Once an order is defined by SuccinctEdge's query optimizer, our system translates TPs into SDS's standard operations: access, rank and select. 
We are using an additional function, namely $rangeSearch(a,b,c)$, which finds all the occurrences of value $c$ in the interval $(a,b)$. It uses a binary search, \ie due to the ordering imposed on portions of WTs,  and returns the indexes of matching values. The use of this function speeds up query execution since it efficiently prunes searches by just computing the boundaries of WTs, \eg $WT_o$ and $WT_s$, instead of scanning all values of an interval.}

We now present two translation examples in Algo. \ref{algo_s} and \ref{algo_o} for resp. the $(s,p,?o)$ and $(?s,p,o)$ TPs.
Algo. \ref{algo_s} shows how to retrieve an answer set with a $(s,p,?o)$ TP. The idea is to first compute an interval of object values related to a given predicate and subject pair. This is performed by navigating through our BM and WT structures. All the objects in this interval are the results of this TP.
Algo. \ref{algo_o} retrieves all the subjects of a $(?s,p,o)$ TP.
Unlike Algo. \ref{algo_s}, we can not locate all the subjects directly, that's why the strategy is to get the interval of all the objects corresponding to the known predicate top-down, after which we locate the object in this interval (there may be multiple appearances) 
and get the corresponding subjects.

\begin{algorithm}
    \caption{Search the triple pattern $(s,p,?o)$}
    \label{algo_s}
    \LinesNumbered
    \KwIn{Predicate s,p}
    \KwOut{Results res}
    \BlankLine
    $id_p\leftarrow FindIdFromDictionary(p)$\;
    $id_s\leftarrow FindIdFromDictionary(s)$\;
    $index_p\leftarrow wt_p.select(1,id_p)$\;
    $index_{sBegin}\leftarrow
    bitmap_{ps}.select(index_p + 1, 1)$\;
    $index_{sEnd}\leftarrow
    bitmap_{ps}.select(index_p + 2, 1)$\;
    \For{$index_s$ in $wt_s.rangeSearch(index_{sBegin},index_{sEnd},id_s$)}{
        $index_{oBegin}\leftarrow bitmap_{so}.select(index_{sBegin} + 1, 1)$\;
        $index_{oEnd}\leftarrow
    bitmap_{so}.select(index_{sEnd} + 2, 1)$\;
        \For{$index_o\leftarrow index_{oBegin}$ \KwTo $index_{oEnd}$}{
            $id_o \leftarrow wt_o[index_o]$\;
            add $(id_s,id_p,id_o)$ into res\;
    }
    }
    return res\;
\end{algorithm}

\oli{In cases where reasoning services are necessary to provide an exhaustive answer set, we can replace $index_p$ with a continuous interval corresponding to a LiteMat interval. This interval is efficiently computed given the order imposed on leaves of a certain WT, \eg $WT_p$ for the property hierarchy. The larger and deeper a property hierarchy, the more efficient is this optimization approach since it prevents from navigating in the complete tree of a given WT.}

TPs containing \texttt{rdf:type} are processed differently using the RDFType store component, where some simple structure look-ups permit to efficiently retrieve to subjects of a given concept or the concepts of a given subject.

The next step corresponds to joining the results obtained from the execution of TPs. This occurs when different TPs share a common variable. One of our joining approach amounts to propagate variable assignments from one TP to another. Consider the triple set of Figure \ref{fig:triple_store_example}(a) and TPs $(?s,p1,o1)$ and $(?s,p2,?o)$. The first TP gets the following assignments:  $?s:\{s1,s2\}$ which will serve to dynamically generate $(s1,p2,?o)$ and $(s2,p2,?o)$ for the second triple.

\begin{figure}
\centering
\includegraphics[height=2.8cm]{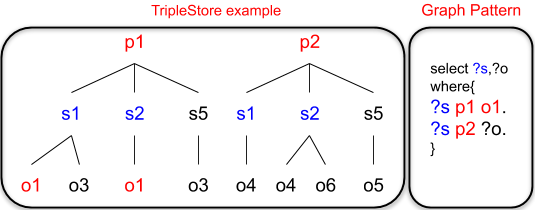}
\caption{Merge join example}
\label{fig:merge_join}
\end{figure}

During the join operation, we can benefit from a merge join (due to the original PSO value order) in certain cases when the values assigned to a joining variable to the TP are kept in order. 
For instance, in the case of a star-shaped basic graph pattern, \eg $(?s,p1,o1)$ and $(?s,p2,?o)$, thanks to the facts that all the subjects connected to a certain predicate are ordered and that all the objects connected to one certain subject are also ordered, we can perform a merge join on the subject variable. 
Figure \ref{fig:merge_join} provides a graph pattern (on the right side) and an RDF Graph (left side). From the first TP, we can retrieve $\{(p1,s1,o1),(p1,s2,o1)\}$ as the answer set. Clearly, since the subjects are ordered for a given predicate, the system can easily use a merge join with the 2nd TP of the query.
In cases where the order is not guaranteed, we use nested loop joins.

\begin{algorithm}
    \caption{Search the triple pattern $(?s,p,o)$}
    \label{algo_o}
    \LinesNumbered
    \KwIn{Predicate p}
    \KwOut{Results res}    \BlankLine
    $id_p\leftarrow FindIdFromDictionary(p)$\;
    $index_p\leftarrow wt_p.select(1,id_p)$\;
    $index_{sBegin}\leftarrow
    bitmap_{ps}.select(index_p + 1, 1)$\;
    $index_{sEnd}\leftarrow
    bitmap_{ps}.select(index_p + 2, 1)$\;
    
    $index_{oBegin}\leftarrow
    bitmap_{so}.select(index_{sBegin} + 1, 1)$\;
    $index_{oEnd}\leftarrow
    bitmap_{so}.select(index_{sEnd} + 2, 1)$\;

    \For{$index_o$ in $wt_o.rangeSearch(index_{oBegin},index_{oEnd},id_o)$}{
        $index_s \leftarrow bitmap_{so}.rank(index_o + 1, 1) - 1$\;
        $id_s \leftarrow wt_s[index_s]$\;
        add $(id_s,id_p,id_o)$ into res\;
    }
    return res\;
\end{algorithm}

Previous executions steps are repeated until all the TPs have been processed. Then the answer set of the query is translated using our dictionaries and presented to the end-user or application.


\section{Related Work}
\label{sec:related}

Header Dictionary Triples (HDT)\footnote{http://www.rdfhdt.org/}\cite{DBLP:conf/esws/Martinez-PrietoGF12} is a compact data structure and binary serialization for RDF data. The Triples component of HDT requires that triples are sorted in a specific order, \eg SPO. The triples are stored in so-called Bitmap Triples which represents a forest of RDF trees, \eg each tree is rooted with a given subject value. The remaining tree layers, \eg for P and O, each correspond to a sequence of identifiers and a bit sequence which connects layers like our bitmaps. Like HDT, SuccinctEdge represents RDF triples as trees but it makes an extensive use of WTs and depends on three different storage approaches, namely Object-triple-store, Datatype-triple-store and RDFType-store. Moreover, SuccinctEdge is equipped with a full-fledge query processing component and supports RDFS reasoning within SPARQL queries.

In \cite{10.5555/3031843.3031879}, a so-called Semantic Index is proposed for Ontology-Based Data Access (OBDA) systems. In this approach, each entity (concept or property) in the corresponding hierarchy is assigned a numerical value according to a breadth-first search traversing of the hierarchy. Provided with this assignment, one is ensured that any sub-hierarchy is associated to a consecutive set of numerical values (\ie an interval). Intuitively, each entity is associated to an interval covering the indices of all its sub-entities. This approach is related to the LiteMat encoding scheme but one benefit of using prefix codes is inherent to the SDS (\ie wavelet tree) used in SuccinctEdge. In a nutshell, answering a request on SuccinctEdge induces the system to visit some data structures. Moreover, Semantic Index is just encoding scheme for a knowledge base and is not a fullfledged query processor.

WaterFowl\cite{waterfowl14} was designed as a first attempt to use SDS for RDF storage and query processing. Although its compactness can be used in an edge computing setting, it lacks the different object storage implementation and query processing (including optimization) features of SuccinctEdge.

RDF4Led is an RDF store designed for edge computing. Compared to our system, RDF4led is disk-based, \ie it store the data on a SD card, and depends on multiple indexes which imply a high memory footprint. Finally, it doesn't support reasoning services nor SPARQL's UNION clause which prevents to apply a query rewriting in order to support reasoning.

ZipG\cite{zipg17} is a distributed graph store designed for the property graph data model. Hence it does not provide support for SPARQL (or any declarative query layer) and reasoning services. Its storage layout is based on the Succinct\cite{succinct15} system and is mainly composed of mostly flat binary unstructured files. ZipG is not compatible with devices located at the edge of a network. In fact, it thrives in a cloud computing setting. 

TerminusDB\footnote{https://terminusdb.com/} is an open-source general-purpose graph database. It aims to store very large graphs in main memory by scaling vertically. Such an approach is not compatible with the edge computing ecosystem that we are targeting. Moreover, TerminusDB does not support Semantic Web standards and hence can not benefit from the existence of a large set of available ontologies to support data integration or support reasoning services associated to RDFS and OWL ontology languages.
\section{Evaluation}
\label{sec:eval}
\subsection{Experimental setting}
Our experimentation is conducted on a Raspberry Pi 3B+ which can be considered as a typical edge computing device on which we can run some sophisticated programs. This small computer is equipped with a Cortex-A53 (ARMv7l) 32-bit SoC 1.4GHz CPU and 1GB LPDDR2 SDRAM. A SD-card, a widely used memory solution on such devices, of 8GB is used as persistent storage.

Considering the evolution of technology, it is widely accepted that edge computing devices will be more and more powerful in the near future. Hence, it is quite obvious that devices with sufficient calculation power and memory, \eg Raspberry Pis, Odroids, etc. , will be deployed in the edge. 

SuccinctEdge is implemented in C++14 and the SDS-lite C++ library\footnote{https://github.com/simongog/sdsl-lite} is required during compilation. More details can be found on github\footnote{https://github.com/xwq610728213/SuccinctEdge}. We are comparing SuccinctEdge against RDF4Led\cite{DBLP:conf/iot/TuanHWP18}, two Apache Jena\footnote{https://jena.apache.org/} (version 3.15) database implementations and RDF4J's Memory Store\footnote{https://rdf4j.org/} (version 3.4.0) . RDF4Led is to the best of our knowledge the only RDF store specifically sdesigned for edge computing. It is characterized by a small memory footprint, although the database system does not reside in the main-memory. The two Apache Jena stores are TDB2 and the in-memory store. They are both open-source  relatively lightweight and  robust RDF  store. RDF4J (originally Sesame) is an open-source Java Framework for managing RDF data. The core RDF4J databases are mainly intended for small to medium-sized datasets and thus it makes sense to consider them for Edge computing. We also considered RDFox \cite{DBLP:conf/semweb/NenovPMHWB15}, a main-memory, centralized RDF store that is designed on a shared-memory architecture, but could not make it work on our raspberry Pi 3B+ since we only had access to a 64-bit pre-compiled version. Systems like Ontotext GraphDB\footnote{https://www.ontotext.com/products/graphdb/}, Stardog\footnote{https://www.stardog.com/}, MarkLogic\footnote{https://www.marklogic.com/}, AllegroGraph\footnote{https://allegrograph.com/}, AWS Neptune\footnote{https://aws.amazon.com/fr/neptune/} have not been considered since they have been designed for massive loads and scalability on high-end servers or Cloud computing.

\subsection{Datasets and queries}
The experimentation uses both synthetic and real-world datasets. This duality is motivated by the current lack of large graphs emitted from sensors at our industrial partner. In fact, our real-world datasets, which correspond to the water management distribution in ENGIE's building, consist of 250 and 500 triples. They are denoted with their number of triples in this experimentation.

Due to these size limitations, it is not possible to stress SuccinctEdge in terms of graph sizes. Hence, we are also experimenting with the synthetic Lehigh University Benchmark (LUBM)\footnote{http://swat.cse.lehigh.edu/projects/lubm/} which can be easily configured to produce large datasets. Starting from a LUBM with one university, \ie composed  of  over  103.000  triples (denoted 100K), we created several triple subsets of 1.000, 5.000, 10.000, 25.000 and 50.000 triples which are respectively denoted as 1K, 5K, 10K, 25K and 50K in the remaining of this section. They are used to evaluate the behaviors of the five evaluated data management systems. Note that some of these synthetic datasets have triple set size way beyond what most sensors are currently emitting in real-world industrial use-cases. All submitted queries are available on the system's Github page. 

\subsection{Experimentation results}
In this section, we are aiming to compare the previously mentioned RDF stores (\ie Jena TDB, Jena-in-memory, RDF4Led, RDF4J and SuccinctEdge) on the following dimensions: graph construction time, memory footprint (\ie the storage space take by different systems with the previous data sets), query execution performances on different triple patterns and basic graph patterns. Lastly, we evaluate the performance (duration time) of queries which necessitate reasoning services to produce an exhaustive answer set.

\subsubsection{Back-end construction time}

The back-end construction time corresponds to the time taken by each system to read the dataset file and to construct its proper storage layout (including indexes in the case of all systems except Succint-Edge which is self-index) on which queries can be asked.

\begin{figure}
\centering
\includegraphics[scale=0.4]{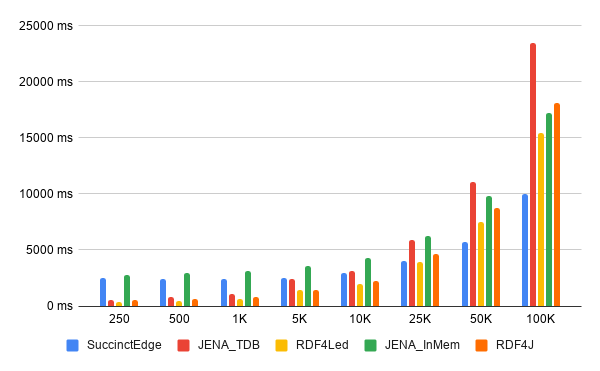}
\caption{Construction time comparison}
\label{fig:construction_time}
\end{figure}

In order to fully evaluate the performances of all the systems, we compare the back-end construction time of these systems with data sets ranging from 250 to over 100.000 triples.
Figure \ref{fig:construction_time} provides details on this experimentation. SuccinctEdge doesn't show much advantage when data set is rather small (up to 1.000 triples). However, as the data set grows larger, our system shows a great advantage compared with the other systems. We attribute this to the fact that the SDS-Lite library which is responsible for creating SuccinctEdge's BMs and WTs has an important start-up overhead that is relatively important compared to the effective duration of the structures. We consider that this may be optimized in future work, but it is out of the scope of this paper.

\subsubsection{Storage size}
As SuccinctEdge is an in-memory RDF system, it is difficult to directly compare the memory occupation with Jena TDB and RDF4Led (which are both disk-based RDF stores). We persisted all the data structures existing in SuccinctEdge to disk in order to make a fair comparison. 



\begin{figure}
\centering
\begin{minipage}[t]{0.48\textwidth}
\centering
\includegraphics[scale=0.35]{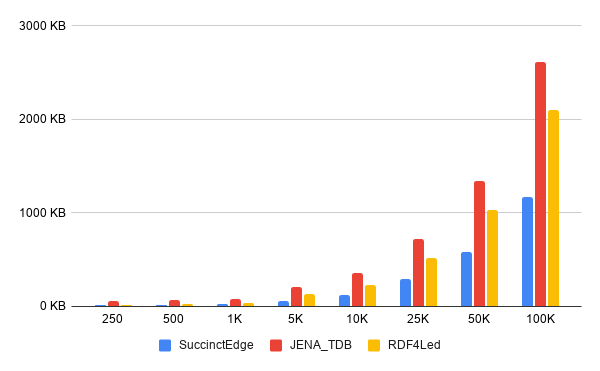}
\caption{Dictionary size comparison}
\label{fig:dict_size}
\end{minipage}
\begin{minipage}[t]{0.48\textwidth}
\centering
\includegraphics[scale=0.35]{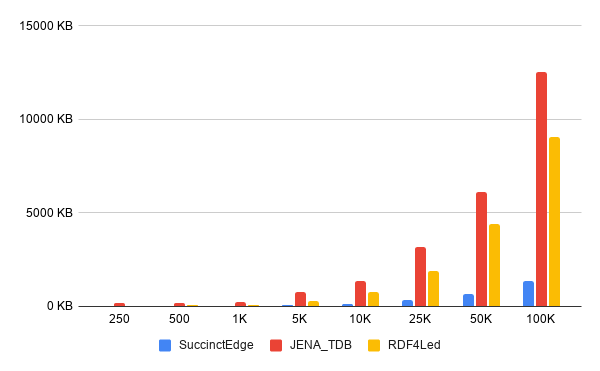}
\caption{Storage size without dictionary comparison}
\label{fig:non_dict_size}
\end{minipage}
\end{figure}

We consider separately the dictionary and triple storage spaces. 
Figure \ref{fig:dict_size} provides the three systems' dictionary sizes for all 8 datasets. In all cases, Jena TDB requires the largest memory footprint and SuccinctEdge takes about half of the size of RDF4Led. 

Considering the triple storage space, displayed in Figure \ref{fig:non_dict_size}, SuccinctEdge consumes much smaller space thanks to its SDS-based storage implementation and self-index approach. This enables to reach one of our goal which is to store as much data as possible in a given amount of RAM.

\begin{figure}
\centering
\includegraphics[scale=0.3]{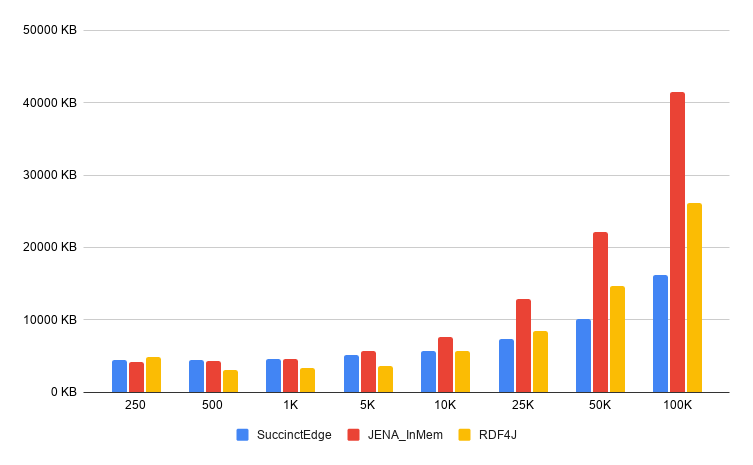}
\caption{RAM footprint comparison}
\label{fig:in_memory_size}
\end{figure}

We are also comparing the main-memory footprint of SuccinctEdge with the in-memory systems, \ie RDF4J and Jena\_InMem. In this evaluation, it is not possible to distinguish between the space used for the dictionaries and the datasets. So we provide the total space amount. Figure \ref{fig:in_memory_size} yields the experiment results. We can see that as the amount of data grows, SuccinctEdge gradually shows its strength in saving memory space. We mainly attribute this to the size of the indexes stored by both RDF4J and Jena\_InMem.

\subsubsection{Triple pattern query}

Considering query processing, we start the evaluation with single triple patterns,\ie excluding the cost of join operations, in order to directly compare the performance of data retrieval in different systems.

We first consider the two interesting triple patterns containing a single variable in the context of SuccinctEdge: \textbf{?s,P,O} and \textbf{S,P,?o}. Moreover, we also consider these two triple patterns with different selectivity, \ie result sets ranging from 4 to 513 tuples.
Table \ref{table:query_one_triple_o_var} and \ref{table:query_one_triple_s_var} provide the results of this experimentation for the LUBM1 dataset (over 100.000 triples).

As said previously, in an IoT setting, we are mainly interested in executing a query on the freshest data and such a query is generally execute only once per graph instance. Hence, we are only considering hot runs. 

SuccinctEdge outperforms other systems on almost all query selectivities. It is only on relatively non-selective, at least considering an IoT context, that SuccinctEdge gets beaten by RDF4J (Table \ref{table:query_one_triple_o_var} for an answer set of 513 tuples and \ref{table:query_one_triple_s_var} for an anwser set of 521 tuples). Considering our potable water distribution running example, the answer set of each query is clearly very selective. That is only a small set of tuples are retrieved from a specific query out of a given measure. We consider that this will be the case for many industrial situations. Thus, high selective queries is clearly an important playground for RDF stores running in edge computing.
In the case of selective queries, SuccinctEdge can be up to one order of magnitude faster than its RDF4J most direct competitor, \eg  Table \ref{table:query_one_triple_o_var} with a result set of size 4).

\begin{table}[]
\caption{Data retrieval for a single \textbf{S,P,?o} TP. The first row represents the number of triples in the answer set. All times in ms. Bold times are a column's most efficient.}
\centering
\begin{tabular}{|c|c|c|c|c|c|}
\hline
& \multicolumn{5}{c|}{\textbf{Query duration}}\\
\hline
\textbf{Systems} & 4 & 66 & 129 & 257 & 513\\
\hline
SuccinctEdge & \textbf{0.3} & \textbf{3.5} & \textbf{6.2} & \textbf{10.9}	 & 23.3\\
\hline
RDF4Led & 12 & 28 & 33 & 47 & 84\\
\hline
Jena\_TDB & 7 & 16 & 22 & 27 & 33\\
\hline
Jena\_InMem & 5 & 11 & 15 & 19 & 29\\
\hline
RDF4J & 3 & 6 & 10 & 11.1 & \textbf{13}\\
\hline
\end{tabular}
\label{table:query_one_triple_o_var}
\end{table}

\begin{table}[]
\caption{Data retrieval for a single \textbf{?s,P,O} TP. The first row represents the number of triples in the answer set. All times in ms. Bold times are a column's most efficient.}
\centering
\begin{tabular}{|c|c|c|c|c|c|}
\hline
& \multicolumn{5}{c|}{\textbf{Query duration}}\\
\hline
\textbf{Systems} & 5 & 17 & 135 & 283 & 521\\
\hline
SuccinctEdge & \textbf{0.7} & \textbf{1.5} & \textbf{10.1} & 20.7 & 32.0\\
\hline
RDF4Led & 6 & 9 & 51 & 71 & 81\\
\hline
Jena\_TDB & 7 & 8 & 30 & 32 & 41\\
\hline
Jena\_InMem & 7 & 8 & 15 & 21 & 27\\
\hline
RDF4J & 3 & 3 & 11 & \textbf{16} & \textbf{21}\\
\hline
\end{tabular}
\label{table:query_one_triple_s_var}
\end{table}

\begin{figure}
\centering
\includegraphics[scale=0.35]{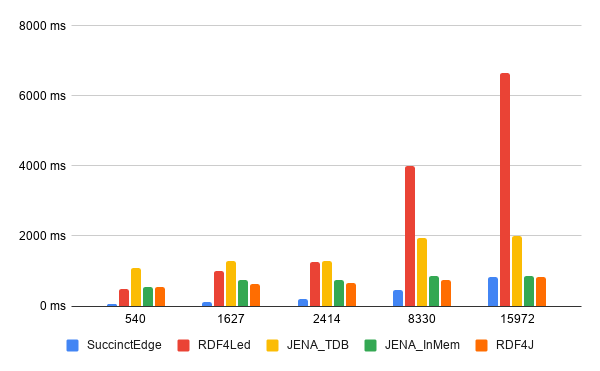}
\caption{Data retrieval of queries with only one triple pattern of type \textbf{?s,P,?o}, the x-axis represents the number of triples in the answer set.}
\label{fig:query_one_triple}
\end{figure}

Figure \ref{fig:query_one_triple} shows the results of several randomly picked \textbf{?s,P,?o} queries (triple patterns with a constant predicate and variable subject and object). We can see from the results that SuccinctEdge outperforms the other systems. 
Clearly, the conclusion obtained on single triple patterns with a single variable that the more selective, the more efficient SuccinctEdge is compared to the other systems, is confirmed. We attribute this to SuccinctEdge's in-memory approach and structure which is \textbf{?s,P,?o}-friendly due to its PSO self-index approach. Moreover Jena TDB and RDF4Led also have PSO or POS indexes but are disk-based database, for whom, loading data from disk takes a non-negligible time. 
The numbers of triples in the answer sets of our single variable TP experimentation are much smaller than that of the \textbf{?s,P,?o}. This leads to greater differences between the different systems. This is again due to the fact that RDF4Led and Jena TDB are loading data from disk. Nevertheless, we can consider that retrieving over 500 tuples at a time from a single sensor is already quite unusual for an IoT use case.
The comparison with in-memory stores (RDF\_InMem and RDJ4) highlights that SuccinctEdge is faster for answer sets lower than 10.000 tuples. At 16.000 result set tuples, The three systems behave similarly. Again, from the point of view of our industrial partner, this is currently unusual for a large portion of industrial use cases.

\subsubsection{Graph pattern query}
\begin{figure}
\centering
\begin{minipage}[t]{0.48\textwidth}
\centering
\includegraphics[scale=0.35]{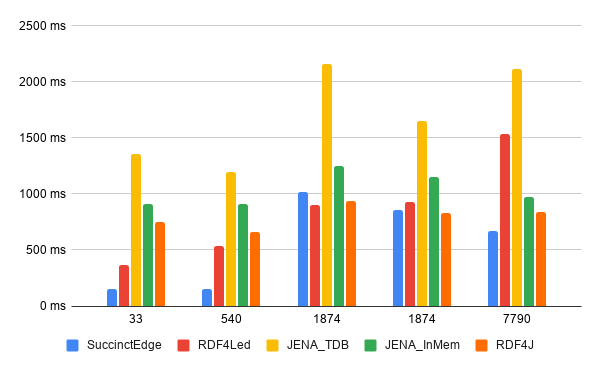}
\caption{Queries with multiple triple pattern (x-axis corresponds to the number of tuples in the answer set)}
\label{fig:query_graph_pattern}
\end{minipage}
\begin{minipage}[t]{0.48\textwidth}
\centering
\includegraphics[scale=0.35]{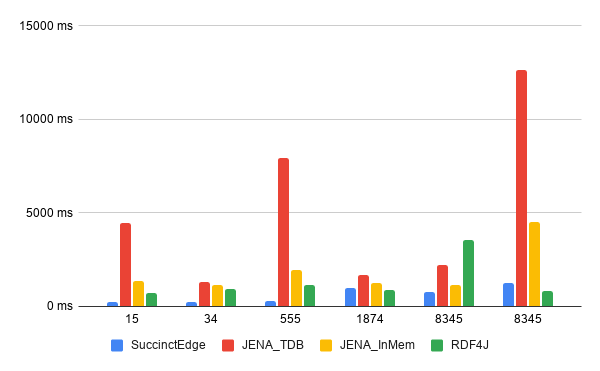}
\caption{Queries with RDFS reasoning (x-axis corresponds to the number of tuples in the answer set)}
\label{fig:query_with_reasoning}
\end{minipage}
\end{figure}

We now compare performances over queries containing multiple triple patterns, \ie requiring joins. 
Four queries with different selectivities (answer sets ranging from 540 to close to 8.000 tuples) have been executed. We can see in Figure \ref{fig:query_graph_pattern} that RDF4Led and SuccinctEdge are always outperforming Jena TDB. SuccinctEdge is either more efficient than RDF4Led or slightly less efficient that RDF4Led. This showcases that in some cases RDF4Led finds a better TP query ordering strategy than SuccinctEdge and/or benefits from its large set of available indexes. Considering the latter, it is a price we are willing to pay for a lower memory footprint. Nevertheless, the former reason emphasizes that we can improve our query optimizer.

The comparison with the in-memory RDF stores emphasizes that the three systems behave similarly except for highly selective queries where SuccinctEdge is again more efficient. The differences between the query executions depend on the patterns used in the BGP of these four queries. Overall, we are satisfied that our system, with a single index, is at least at the same level than the two other systems.


\subsubsection{Query with RDFS reasoning}
Our final experimentation concerns queries requiring some reasoning services. We have generated several queries containing a mixture of RDFS:subClassOf and RDFS:subPropertyOf inferences. These queries present different selectivity characteristics, ranging from 15 to 8.345 tuples in the answer sets.

For SuccinctEdge, the reasoning service is automatically supported by LiteMat's encoding and is hence native in the system. This is not the case for the  other systems for which we have rewritten each query as the union of all the possible sub-queries. Since RDF4Led doesn't support the SPARQL UNION clause no results are presented in Figure \ref{fig:query_with_reasoning} for this system. Obviously, SuccinctEdge is much more efficient than Jena TDB. It is quite logical that the more entailments the query requires, the more efficient SuccinctEdge is compared to a system like Jena TDB, \eg with the last three queries, the query time is nearly ignorable with SuccinctEdge compared to Jena TDB.

As for Jena\_InMem, it performs better than Jena\_TDB while still falling behind SuccinctEdge. When compared with RDF4J, SuccinctEdge performs better or similarly depending on the complexing of the reasoning services, \ie number of SPARQL UNION clauses. Note that we provide manual query rewriting to the Jena and RDF4J systems while these systems could implement the reasoning task with their APIs. In doing some, we provide a clear advantage to these systems since they do not have to load the ontology to perform the reasoning. Moreover, the extra cost of computing the UNION rewriting is not considered in the times of the Jenas and RDF4J executions.


\section{Conclusion}
\label{sec:conclusion}
We have presented the first, to the best of our knowledge, KG inference-enabled data management system design for edge computing. Due to its unique index, compact, in-memory approach, we have demonstrated that SuccinctEdge outperforms its direct competitors on the following dimensions: query performance on different query patterns, efficiency of reasoning services, back-end size and creation time.  The system is currently being deployed at some large building facilities at ENGIE and should help in detecting anomalies in water distribution and energy consumption. Due to its generic nature, SuccinctEdge is relevant for many IoT use cases such as anomaly and risk detection, supervising energy production and distribution. In the future, we are aiming to improve the query optimizer and support queries ranging several graphs. We are also considering to design a more efficient management of objects linked to datatype properties and to increase the expressiveness of supported ontology languages, \eg RDFS++ and OWL2RL. Moreover, we are considering the possibility of exchanging information with a larger graph portion that would reside in a cloud store.

\bibliographystyle{ACM-Reference-Format}
\bibliography{main}

\appendix
\section{Queries}
\label{app:queries}
\begin{table*}
\caption{Query summary with the following notations: 'SS' and 'OS' respectively correspond to subject, subject and object,subject joins; 'Co' for concept hierarchy inferences, 'Pr' for property hierarchy inferences}
\centering
\begin{tabular}{|c|c|c|c|c|c|c|c|c|c|c|c|c|c|c|}
\hline
& \multicolumn{14}{c|}{\textbf{Queries}}\\
\hline
\textbf{Systems} & S1-5 & S6-10 & S11-15 & M1 & M2 & M3 & M4 & M5 & R1 & R2 & R3 & R4 & R5 & R6\\
\hline
TP number & 1 & 1 & 1 & 2& 3& 5& 3& 11 &5& 5& 3& 6& 3 & 11\\
\hline
TP type(s) & sp? & ?po & ?p? & ?p? & ?p? & ?p? & ?p? & ?p? & ?p? & ?p? & ?p? & ?p? & ?p? & ?p?\\
 & & & & & ?po & ?po & ?po & ?po & ?po & ?spo & ?po & ?po & & ?spo\\
 & & & & & & & sp?o & & & & & & & sp? \\
 \hline
Join type & - & - & - & SS & SS & SS,OS & OS & SS,OS & SS,OS & SS,OS & SS &SS,OS & OS& SS,OS\\
         &   &   &    &    &    &       &    & OO    &       &       &    &     &      & OO\\
\hline
Join number & 0 & 0 & 0 & 1 & 2 & 4 & 4& 10&4& 2 & 2& 5& 2 & 10\\
\hline
Path length & 1 & 1 & 1 & 1 & 1 & 3 & 3& 4&3& 3 & 1& 3 & 3 & 4\\
\hline
Selectivity & [4,513] & [5,521] & [540,15972] & 540 & 1874 & 1874& 7790 & 33 & 15 & 555& 1874& 1874& 8345& 34\\
\hline
Derived & 0 & 0 & 0 &0& 0& 0& 0& 0& 15 & 540 & 1874 & 1874 & 555& 1\\
triples &  &  &  && & & & &  & &  &  & &\\
\hline
Reasoning & - & - & - & - & -& -& - & - & Co & Co & Co & Co & Pr & Pr\\
type &  &  &  &  & & &  &  &  & Pr & Pr & Pr & & \\
\hline
\end{tabular}
\label{table:queryMetrics}
\end{table*}
A total of 26 queries have been evaluated over a LUBM dataset consisting of over 100.000 triples. They can be dispatched into 2 groups: whether their contain a single triple pattern or multiple ones. In this section, we list only the most prominent queries and provide templates for the other ones. Moreover, we present their main characteristics. The interested reader can access all of them on the paper companion GitHub page\footnote{https://github.com/xwq610728213/SuccinctEdge}. The following prefixes apply to all queries: lubm <http://swat.cse.lehigh.edu/onto/univ-bench.owl\#>, rdf <http://www.w3.org/1999/02/22-rdf-syntax-ns\#> 

\subsection{Single triple pattern queries}
This first set of queries contain a single triple pattern in the WHERE clause. We distinguish between queries with a single variable ,either at the object (denoted sp?) or subject (denoted ?po) position, from queries with two variables (denoted ?p?). As explained in the paper, we do not consider that variables at the property position make sense in SuccinctEdge's use cases.

 \subsubsection{SP?o queries}
The identification of these 5 queries range from S1 to S5. We used the following query template:
 \begin{verbatim}
SELECT ?X WHERE {X1 P1 ?X} \end{verbatim}
For S1, P1 binds to the lubm:takesCourse properperty and X1 is an undergraduate student constant. For queries S2 to S5, P1 binds to lubm:publicationAuthor and the X1 bind to different publication instances. The cardinality of these queries are in Table \ref{table:query_one_triple_o_var}.

 \subsubsection{?sPO queries}
These queries are identified from S6 to S10 and correspond to the following query template:
\begin{verbatim}
SELECT ?X WHERE { ?X P1 O1 }
\end{verbatim}
P1 and O1 correspond to property and individual constants which for S6 to S10 respectively take the values (all properties are in the lubm namespace) : advisor/assistant professor constant, takesCourse/ course constant, worksFor/department constant, name/ publication constant, memberOf/ department constant.


\subsubsection{?sP?o queries}
\begin{verbatim}
S11: SELECT ?X ?Y ?Z WHERE { ?X lubm:worksFor ?Z }
S12: SELECT ?X ?Y ?Z WHERE { ?X lubm:teacherOf ?Y}
S13: SELECT ?X ?Y ?Z WHERE { 
     ?X lubm:undergraduateDegreeFrom ?Y .}
S14: SELECT ?X ?Y ?Z WHERE { ?X lubm:emailAddress ?Y }
S15: SELECT ?X ?Y ?Z WHERE { ?X lubm:name ?Y }
\end{verbatim}
 
\subsection{Multiple triple patterns queries}
In this set of queries, the BGP is composed of several triple patterns. The 11 queries in this category can be decomposed into those requiring or not some reasoning form (either based on concept or property hierarchies).

 \subsubsection{No inference queries}
\begin{verbatim}
M1: SELECT ?X ?Y ?Z WHERE { ?X lubm:worksFor ?Z .
 ?X lubm:name ?Y .}
M2: SELECT ?X ?Y ?Z WHERE { ?X lubm:memberOf ?Z .
 ?X rdf:type lubm:GraduateStudent .
 ?X lubm:undergraduateDegreeFrom ?Y .}
M3:  SELECT ?X ?Y ?Z WHERE { ?X lubm:memberOf ?Z . 
 ?X rdf:type lubm:GraduateStudent .
 ?Z rdf:type lubm:Department . 
 ?Z lubm:subOrganizationOf ?Y .
 ?Y rdf:type lubm:University .}    
M4: SELECT ?X ?Y ?Z WHERE { ?X lubm:memberOf ?Z .
 ?Z lubm:subOrganizationOf ?Y .
 ?Y rdf:type lubm:University }
M5: SELECT * WHERE {
 <http://www.Department0...Publication14> 
 lubm:publicationAuthor ?p. ?st lubm:memberOf ?o2.
 ?p a lubm:AssociateProfessor. ?p lubm:worksFor ?o.
 ?o a lubm:department. ?o lubm:subOrganizationOf ?u.
 ?u a lubm:University. ?p lubm:teacherOf ?te.
 ?te a lubm:Course. ?st lubm:takesCourse ?te.
 ?st a lubm:UndergraduateStudent. }    
\end{verbatim}
 \subsubsection{Inference queries}
The identifier of these queries is prefixed if an 'R' since they involve a form of reasoning. The inferences mainly concern 
\begin{verbatim}
R1: SELECT ?X ?Y ?Z WHERE { ?X rdf:type lubm:Person .
 ?Z rdf:type lubm:Department . ?X lubm:headOf ?Z . 
 ?Z lubm:subOrganizationOf ?Y .
 ?Y rdf:type lubm:University .}
R2: SELECT ?X ?Y ?Z WHERE { ?X rdf:type lubm:Person .
 ?Z rdf:type lubm:Department .  ?X lubm:worksFor ?Z . 
 ?Z lubm:subOrganizationOf ?Y .
 ?Y rdf:type lubm:University .}
R3: SELECT ?X ?Y ?Z WHERE { ?X lubm:memberOf ?Z .
 ?X rdf:type lubm:Student .
 ?X lubm:undergraduateDegreeFrom ?Y .}
R4: SELECT ?X ?Y ?Z ?N WHERE { ?X rdf:type lubm:Person .
 ?Z rdf:type lubm:Department . ?X lubm:memberOf ?Z . 
 ?Z lubm:subOrganizationOf ?Y .  ?Y lubm:name ?N.
 ?Y rdf:type lubm:University . }
R5: identical to M4 but reasons over the memberOf property
R6: identical to M5 but reasons over the memberOf and 
worksFor properties.    
\end{verbatim}

\end{document}